\documentclass[twocolumn, times]{aastex631}

\usepackage{booktabs}
\usepackage{multirow}
\usepackage{amsmath}
\hypersetup{colorlinks=true, citecolor=blue, urlcolor=blue, linkcolor=blue, allcolors = blue}

\begin{document}

\title{COSMOS2025: A Machine Learning Census of Massive Quiescent Galaxies at $2.5 < \text{z} < 5$}


\author{Vahid Asadi~\href{https://orcid.org/0009-0005-8897-2385}{\includegraphics[scale=0.04]{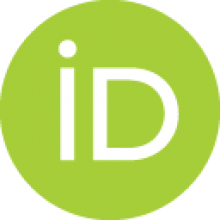}}}
\affiliation{Department of Physics, Institute for Advanced Studies in Basic Sciences (IASBS), PO Box 11365-9161, Zanjan, Iran; \url{vahijd.asadij@gmail.com}}

\author{Hosein Haghi~\href{https://orcid.org/0000-0002-9058-9677}{\includegraphics[scale=0.04]{orcid-ID.png}}}
\affiliation{Department of Physics, Institute for Advanced Studies in Basic Sciences (IASBS), PO Box 11365-9161, Zanjan, Iran; \url{vahijd.asadij@gmail.com}}
\affiliation{Helmholtz-Institut f\"ur Strahlen-und Kernphysik (HISKP), Universit\"at Bonn, Nussallee 14-16, D-53115 Bonn, Germany}
\affiliation{School of Astronomy, Institute for Research in Fundamental Sciences (IPM), PO Box 19395 - 5531, Tehran, Iran}

\author{Akram Hasani Zonoozi~\href{https://orcid.org/0000-0002-0322-9957}{\includegraphics[scale=0.04]{orcid-ID.png}}}
\affiliation{Department of Physics, Institute for Advanced Studies in Basic Sciences (IASBS), PO Box 11365-9161, Zanjan, Iran; \url{vahijd.asadij@gmail.com}}
\affiliation{Helmholtz-Institut f\"ur Strahlen-und Kernphysik (HISKP), Universit\"at Bonn, Nussallee 14-16, D-53115 Bonn, Germany}


\begin{abstract}
The existence of massive quiescent galaxies at high redshifts ($ \text{z} \gtrsim 2$) strongly constrains the rapid quenching mechanisms in galaxy evolution models. We present a machine learning framework to identify massive
($\log(\text{M}_*/\text{M}_\odot) > 9.5$) quiescent galaxies at $2.5 < \text{z} < 5$ in the COSMOS2025 catalog. We train a \texttt{CatBoostClassifier} on mock photometry from the Santa Cruz semi-analytic models (SAMs), incorporating key JWST NIRCam bands and realistic noise to transfer the SAM-derived quiescent label (based on specific star-formation rate) to the observational space. When validated against the SAM ground truth, our classifier achieves a significantly higher recall (completeness) of 78\% (compared to 53\% for spectral energy distribution (SED) fitting), while maintaining a high purity of 82\%. Applied to the COSMOS2025 sample, and assuming the SAM definition of quiescence transfers to the real Universe, the model identifies 1,111 quiescent candidates, a population 2.6 times larger than the 427 candidates identified via the catalog's simple SED‑fitting configuration. Under the SAM definition of quiescence, this consistent pattern of high purity but poor completeness suggests that the SED-fitting methods, constrained by simplified parametric star-formation histories, may miss a significant fraction of the quiescent population, likely galaxies in crucial transitional evolutionary stages. The trained classifier and classified COSMOS2025 sample are publicly available.
\end{abstract}


\section{Introduction}
Massive quiescent galaxies represent crucial laboratories for understanding galaxy evolution and the physical processes that drive the cessation of star formation. The existence of this population at high redshifts ($\text{z} \gtrsim 2$) has been firmly established over the past decade \citep{glazebrook2017massive, schreiber2018near, girelli2019massive, santini2021emergence, carnall2023surprising, nanayakkara2024population, kakimoto2024massive}, with their number densities providing strong constraints on feedback processes and quenching mechanisms in galaxy formation models \citep{somerville2015star,beckmann2017cosmic,merlin2019red,donnari2021quenched,baker2025abundance}.

Multiple quenching mechanisms have been proposed to explain the rapid cessation of star formation in these early massive systems. These include Active Galactic Nucleus (AGN) feedback \citep{croton2006many,wellons2015formation}, stellar feedback \citep [e.g, ][]{hopkins2010maximum,grudic2019elephant}, and merger-induced processes \citep [e.g, ][]{wellons2015formation,rodriguez2019mergers}. At high redshifts, the young age of the universe necessitates rapid and powerful quenching mechanisms, with AGN feedback emerging as a leading candidate \citep [e.g, ][]{davies2024jwst,belli2024star}.

Traditional methods for identifying quiescent galaxies rely on color-color selection techniques, particularly the UVJ diagram \citep{strateva2001color, williams2009detection, van20143d, shahidi2020selection}. The rest-frame NUVrJ diagram is also widely employed as an alternative \citep{ilbert2013mass, davidzon2017cosmos, weaver2023cosmos2020, shuntov2026mass}, as the NUV band is more sensitive to short-timescale variations in star formation ($\sim 10$--$100$ Myr) than the U band ($\sim 100$--$300$ Myr) \citep{arnouts2013encoding}. While widely used, these methods simplify the rich spectral energy distribution (SED) information into a few color indices, potentially missing valuable information and introducing selection biases. As a result, galaxies identified through color selection typically require follow-up SED-fitting \citep{wiklind2008population, girelli2019massive} or spectroscopic confirmation \citep{glazebrook2017massive, schreiber2018near}, which can be time-consuming and resource-intensive for large samples.

\begin{figure*}
	\centering
	\includegraphics[width=0.91\linewidth]{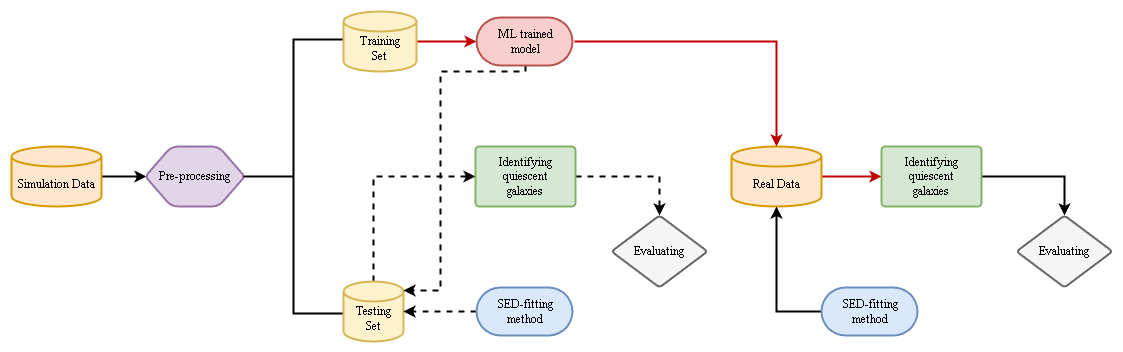}
	\caption{Schematic of the ML pipeline for identifying quiescent galaxies.}
	\label{fig:fig0}
\end{figure*}

Machine learning (ML) approaches offer a promising alternative by leveraging the full multi-wavelength information to identify complex patterns in galaxy SEDs that may be missed by simple color criteria \citep{steinhardt2020method, humphrey2023euclid}. The success of these methods is well demonstrated by \cite{Asadi_2025}, who classified galaxies in the COSMOS2020 catalog into quiescent and star-forming populations using a \texttt{CatBoostClassifier} \citep{prokhorenkova2018catboost} trained on Santa Cruz semi-analytic models (SAMs) \citep{somerville2021mock,yung2022semi}. Their approach, which utilized 28 color indices across the redshift range $0.2 < z < 3.5$, showed that ML methods can significantly outperform traditional techniques by improving especially completeness in quiescent galaxy selection. Furthermore, the computational efficiency of ML algorithms enables the rapid processing of large datasets essential for current and next-generation surveys \citep[e.g, ][]{Hemmati2019,Davidzon2022,asadi2025leveraging}. However, the success of supervised ML approaches depends critically on the quality and representativeness of the training data.

In this study, we present a ML framework for identifying massive quiescent galaxies at $2.5 < \text{z} < 5$ in the COSMOS2025 catalog \citep{shuntov2025cosmos2025}. Building on our previous work \citep{Asadi_2025}, we train a \texttt{CatBoostClassifier} on physically-motivated mock observations from the updated James Webb Space Telescope \citep[JWST; ][]{gardner2006james} version of the Santa Cruz SAMs \citep{somerville2021mock, yung2022semi}, now incorporating 4 JWST NIRCam (F115W, F150W, F277W, F444W) bands that are particularly valuable for hunting high-redshift quiescent galaxies. We employ realistic observational noise, as in our previous study, to bridge the gap between simulation and observation, enabling a complete ML census of the high-redshift quiescent population in the largest contiguous JWST deep field to date.

This paper is organized as follows: Section~\ref{sec:2} describes the simulation and observational datasets (A schematic overview of our complete analysis pipeline is presented in Figure~\ref{fig:fig0}); Section~\ref{sec:3} details our pre-processing methodology; Section~\ref{sec:4} presents our ML classification framework and comparison with SED-fitting; Section~\ref{sec:5} presents our results; and Sections~\ref{sec:6} and~\ref{sec:7} provide discussion and conclusions, respectively.

We adopt a flat $\Lambda$CDM cosmology with parameters $\text{H}_{0}=70\text{kms}^{-1}\text{Mpc}^{-1}$, $\Omega_{\text{m}}=0.3$ and $\Omega_{\Lambda}=0.7$. All magnitudes are reported in the AB system \citep{oke1983secondary}.

\section{Data} \label{sec:2}
This study leverages two complementary datasets: a sophisticated SAM simulation that provides a physically-motivated training set with known galaxy properties, and the state-of-the-art COSMOS2025 observational catalog \citep{shuntov2025cosmos2025} that serves as our target for identifying high-redshift massive quiescent galaxies. The following subsections detail both datasets.

\subsection{Simulation Galaxy Catalog} \label{sec:2:1}
For this study, we utilized JWST wide-field light cones generated from an updated Santa Cruz SAM framework \citep{somerville2021mock,yung2022semi}. This computational approach models the complex physics of galaxy evolution within the context of dark matter halo growth. The model is built upon dark matter merger trees derived from the high-resolution Bolshoi-Planck simulation \citep{klypin2011dark}, which follows a $\Lambda$CDM cosmology. By applying parameterized prescriptions for key processes—including gas accretion and cooling, star formation, supernova feedback, and galaxy mergers—the SAM predicts a comprehensive suite of galaxy properties such as stellar mass, star formation rate, and metallicity.

To emulate the conditions of deep-field surveys, the galaxy data are structured into light cones that represent the three-dimensional spatial distribution across cosmic time. The dataset encompasses five independent fields with geometries overlapping the five Cosmic Assembly Near-IR Deep Extragalactic Legacy Survey (CANDELS) legacy fields \citep{grogin2011candels,koekemoer2011candels}: GOODS-S, GOODS-N, COSMOS, EGS, and UDS. Each field features eight realizations covering redshifts from $z=0$ to $z=10$. This catalog supplies observed-frame photometry for a wide array of observatories, including JWST (NIRCam) \citep{gardner2006james}, Roman (WFI) \citep{spergel2015wide}, Hubble (WFC3/ACS) \citep{dressel2012wide}, Spitzer \citep{werner2004spitzer}, Euclid \citep{laureijs2011euclid}, Rubin \citep{ivezic2019lsst}, GALEX \citep{martin2005galaxy}, SDSS \citep{york2000sloan}, UKIRT \citep{lawrence2007ukirt}, VISTA \citep{emerson2006visible}, and DECam \citep{flaugher2015dark}. It also provides rest-frame luminosities for NUV, FUV, and Johnson/Bessel/Cousins bands, alongside key physical properties like mass, star formation rate, and metallicity for halos and galaxies.

Within this framework, galaxies are populated into dark matter halos, with the most massive galaxy designated as the central and the others as satellites. Satellite galaxies can experience significant mass loss from tidal stripping and dynamical friction, potentially leading to destruction before merging with the central galaxy \citep{somerville2008semi}. Infalling gas is shock-heated and then cools according to a spherical density profile \citep{white1991galaxy}, with accretion proceeding via hot and cold modes that collectively govern the fuel available for star formation.

Star formation proceeds through two primary channels: a quiescent mode based on an updated Kennicutt-Schmidt relation tied to $\mathrm{\text{H}_{2}}$ surface density \citep{somerville2015star}, and a starburst mode triggered by rapid gas compression during galaxy mergers, modeled on binary merger simulations \citep{robertson2006fundamental, hopkins2009disks}. Feedback from massive stars and supernovae is a critical component, as it can eject cold gas from galaxies and thereby regulate their evolution \citep{somerville2008semi}. Furthermore, each galaxy hosts a central black hole seed that can grow through a quasar mode, driven by mergers and instabilities, and a radio mode, fueled by the accretion of hot halo gas as described by the Bondi-Hoyle-Lyttleton model \citep{bondi1952spherically}.

The parameters governing these physical processes are rigorously calibrated against observational data, such as galaxy mass functions, luminosity functions, and scaling relations like the Tully-Fisher and black hole–bulge mass relations. This calibration ensures that the SEDs generated by the SAM are consistent with real observations. These SEDs are subsequently processed to include the effects of dust attenuation and intergalactic medium absorption \citep{madau1995radiative}, and are finally convolved with filter response functions to produce realistic broadband fluxes and magnitudes \citep{yung2019semi, somerville2021mock, yung2022semi}.

\begin{figure}
	\centering
	\includegraphics[width=\linewidth]{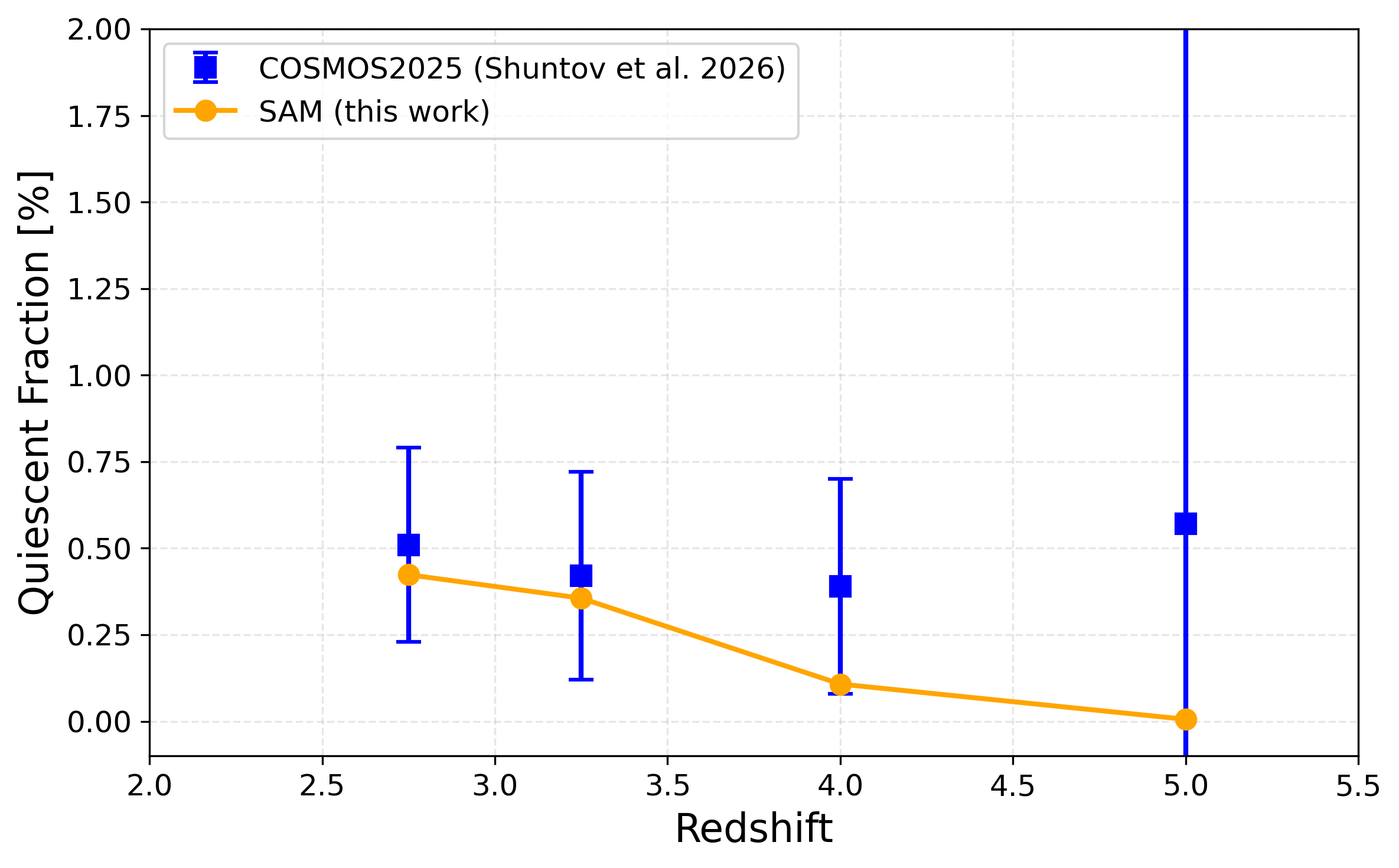}
	\caption{Comparison of the quiescent fraction among massive galaxies between the SAM (sSFR selection ((Equation~\ref{eq1}), orange) and the NUVrJ‑selected observations from COSMOS2025 \citep{shuntov2026mass} (blue squares).}
	\label{fig:fig1}
\end{figure}

To verify that the SAM provides a realistic training set, we compare the fraction of massive quiescent galaxies in the SAM with the observed NUVrJ quiescent fractions from COSMOS2025 \citep{shuntov2026mass}. Adopting the same stellar mass limits and redshift bins as that study, we find that the SAM sSFR‑based quiescent fractions (see Equation~\ref{eq1}) are consistent with the observations within the reported uncertainties across the full redshift range $2.5<\text{z}<5.5$ (Figure~\ref{fig:fig1}). The decline of the quiescent fraction with increasing redshift is well reproduced. This agreement confirms that the SAM can provide a realistic description of the quiescent population, justifying its use as a training set.

\subsection{Observational Galaxy Catalog} \label{sec:2:2}
The analysis in this work is based on the COSMOS2025 galaxy catalog \citep{shuntov2025cosmos2025}, the definitive data release from the COSMOS-Web JWST Treasury program (GO\#1727; PIs: Casey \& Kartaltepe). COSMOS2025 represents a major advancement over previous COSMOS catalogs \citep{laigle2016cosmos2015,weaver2022cosmos2020}, providing photometry, morphology, photometric redshifts, and physical parameters for over 700,000 galaxies in the central $\sim 0.54 \, \rm{\text{deg}}^2$ of the Cosmic Evolution Survey (COSMOS) field \citep{scoville2007cosmic}.

The catalog combines deep, high-resolution imaging from space- and ground-based observatories. This includes JWST/NIRCam (F115W, F150W, F277W, F444W) and JWST/MIRI (F770W) data from COSMOS-Web, complemented by HST/ACS F814W imaging. Ground-based coverage includes ultraviolet (CFHT/MegaCam), optical (Subaru/HSC and Suprime-Cam), and near-infrared \citep[VISTA/VIRCAM; UltraVISTA DR6; ][]{mccracken2012ultravista} data.

Source detection is performed on a point-spread function (PSF)-homogenized $\chi^2$ coaddition of the four NIRCam bands using a dual-extraction method to ensure high completeness and purity. The catalog provides aperture photometry on PSF-homogenized images and total model photometry for all 37 bands, derived using a profile-fitting technique with \textsc{SExtractor++}. This method fits Sérsic and bulge+disk models directly to the native-resolution images, convolved with their respective PSFs, providing robust total fluxes and consistent colors across the different datasets.

The rich 37-band photometric coverage ($0.3$--$8\, {\micron}$) enables high-accuracy SED-fitting. Photometric redshifts and physical parameters are derived using \texttt{LePhare} \citep{arnouts1999measuring,ilbert2006accurate} and \texttt{CIGALE} \citep{boquien2019cigale}. The photometric redshifts achieve $\sigma_{\rm \text{NMAD}} = 0.012$ for $m_{\rm \text{F444W}} < 28$, remaining at $\sigma_{\rm \text{NMAD}} \lesssim 0.03$ as a function of magnitude, color, and galaxy type---a factor of $\sim 2$ improvement over COSMOS2020 \citep{weaver2022cosmos2020} at 26 AB mag. The deep NIRCam data also enhance stellar mass completeness; the catalog is $\sim 80\%$ complete at $\log(\text{M}_{\star}/{\rm \text{M}}_{\odot}) \sim 9$ at $\text{z} \sim 10$ and at $\log(\text{M}_{\star}/{\rm \text{M}}_{\odot}) \sim 7$ at $\text{z} \sim 0.2$, an improvement of $\sim 1$ dex over COSMOS2020.

For this study, we use the \textsc{SExtractor++} model-fit photometry, and specifically adopt the associated photometric redshifts and derived physical parameters from the \texttt{LePhare} code. The catalog's superior depth, robust stellar mass completeness at high redshift, and excellent photometric redshift accuracy are critical for conducting a complete and reliable census of high-redshift massive quiescent galaxies, which is the primary goal of this work.

\section{Pre-processing} \label{sec:3}
The core methodology of this work involves training a ML model on simulated data from the SAM and subsequently applying it to the observational COSMOS2025 catalog to identify high-redshift massive quiescent galaxies. A critical prerequisite for the success of this approach is the careful pre-processing of both datasets. The key principle is to realistically adjust the SAM data to resemble the COSMOS2025 data as closely as possible. This ensures that the training set is a representative proxy for the target domain, which is essential for the model to achieve robust and reliable performance when applied to real observations.

\begin{table}[t]
	\centering
	\caption{Overview of the eleven COSMOS2025 selected bands.}
	\setlength{\belowcaptionskip}{10pt}
	\begin{tabular}{ccccc}
		\hline \hline
		
		Instrument                & Band  & Central\footnote{Median of the transmission curve.}                 & Width\footnote{Full width of the transmission curve at half maximum.}     & Depth\footnote{5$\sigma$ depth computed in empty apertures with diameters of $1.0^{\prime\prime}$ for the ground-based, $0.15^{\prime\prime}$ for the space-based JWST/NIRCam and HST/ACS and $0.5^{\prime\prime}$ for JWST/MIRI images, averaged over the NIRCam area.} \\
		/Telescope                &         &$\lambda$[{\AA}]          & [{\AA}]   &        \\
		(Survey)                  &         &                         &           &         \\
		\midrule
		NIRCam       & F115W        & 11622                  & 2646       & 27.2                \\
		& F150W        & 15106                  & 3348       & 27.4                \\
		& F277W        & 28001                  & 6999       & 28.1                \\
		& F444W        & 44366                  & 11109       & 28.0                \\
		\midrule
		ACS/HST            & F814W    & 8333                  & 2511      & 27.5                     \\
		\midrule
		VIRCAM             & Y        & 10216                 & 923       & 25.8              \\
		/VISTA             & J        & 12525                 & 1718      & 25.8              \\
		UltraVISTA         & H        & 16466                 & 2905      & 25.5              \\
		DR6                & $\text{K}_{s}$  & 21557                 & 3074      & 25.3              \\
		\midrule
		IRAC               & ch1     & 35686             & 7443      & 26.4               \\
		/\textit{Spitzer}  & ch2     & 45067             & 10119     & 26.3               \\
		\hline
		\label{tab:tab1}
	\end{tabular}
\end{table}

\subsection{Selecting Mutual Bands} \label{sec:3:1}
We selected eleven mutual photometric bands between the SAM and COSMOS2025 catalogs: the NIRCam F115W, F150W, F277W, and F444W bands; the ACS F814W band; the UVISTA Y, J, H, and $\text{K}_s$ bands; and the two $Spitzer$/IRAC bands (ch1, ch2). These bands are detailed in Table~\ref{tab:tab1}.

\subsection{Constructing Colors} \label{sec:3:2}
Following the mutual band selection, we constructed colors from the eleven photometric bands to serve as the primary features for our ML model. To maximize the informational content and allow the ML model to learn complex non-linear SED features, we generated every possible unique color combination, resulting in a total of 55 distinct colors. This includes both short-baseline colors (e.g., F150W$-$F277W, J$-$H) and long-baseline colors (e.g., F814W$-$F444W, $\text{K}_s$$ - $ch1).

We opted to use color indices rather than apparent magnitudes because they provide a more direct and robust probe of the physical properties of galaxies. Unlike apparent magnitudes, which are sensitive to distance-dependent effects and flux calibration uncertainties, color indices are distance-independent and more resilient to systematic biases. Additionally, because colors measure the relative flux across different wavelengths, they better constrain the shape of the SED, which is crucial for distinguishing between different stellar populations and dust attenuation effects.

\subsection{Sample Selection} \label{sec:3:3}
To identify high-redshift massive quiescent galaxies, we applied three selection criteria to both the SAM and COSMOS2025 catalogs.

\begin{figure}
    \centering
    \includegraphics[width=\linewidth]{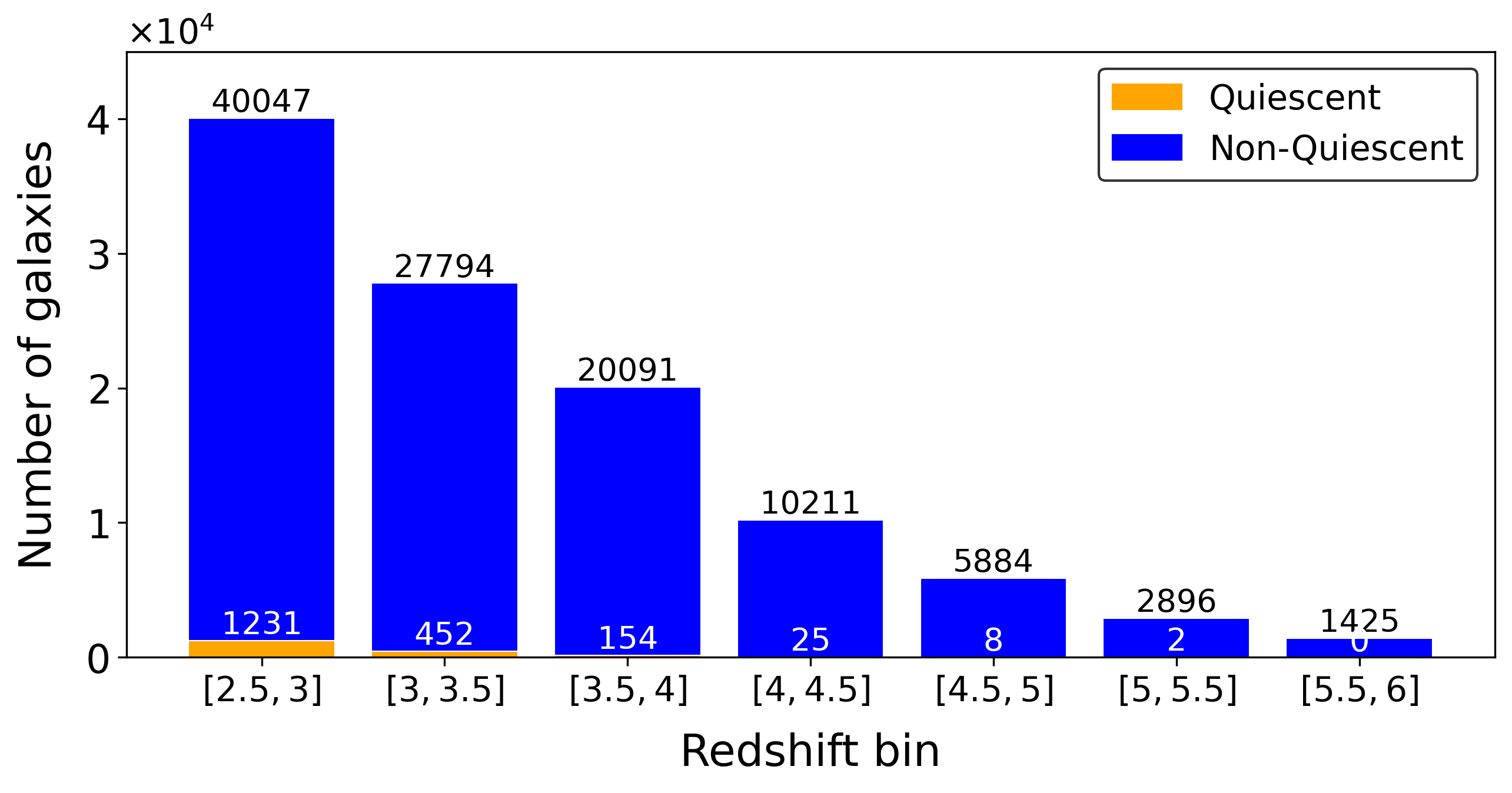}
    \caption{The redshift distribution of quiescent and non-quiescent galaxies in the SAM sample used for model training. The negligible fraction of quiescent galaxies at $\text{z} > 5$ demonstrates the necessity of the imposed redshift cut to ensure a reliable and robust model.}
    \label{fig:fig2}
\end{figure}

\begin{itemize}
    \item \textit{Magnitude cut}: $\text{F444W} < 28$. This limit is chosen based on the exceptional photometric redshift accuracy of the COSMOS2025 catalog (Section~\ref{sec:2:2}).

    \item \textit{Redshift range}: $2.5 < \text{z} < 5$. The lower limit of $\text{z}=2.5$ focuses our search on the high-redshift universe. The upper limit of $\text{z}=5$ is imposed because, as shown in Figure~\ref{fig:fig2}, the number of quiescent galaxies identified in the SAM catalog becomes negligible beyond this redshift (Section~\ref{sec:3:4}). Including these high-redshift populations is impractical due to the insufficient number of quiescent galaxies, which prevents a ML model from learning a robust and reliable identification rule.

    \item \textit{Stellar mass cut}: $\log(\text{M}_* / \text{M}_{\odot}) > 9.5$. This selects for massive galaxies, which is the primary population of interest for studying early quiescent systems.
\end{itemize}

For the SAM catalog, the redshift and stellar mass values are taken directly from the simulation outputs. For the COSMOS2025 catalog, we utilized the photometric redshifts and stellar mass estimates derived from SED-fitting with the \textit{LePhare} code, as provided in the catalog. Additionally, for the COSMOS2025 sample, we applied a further quality cut by limiting the multiplicative error (flux error / flux) to be less than 5 in all selected bands (Table~\ref{tab:tab1}) to exclude objects with excessively noisy photometry.

After applying these criteria, our final sample consists of 105,876 galaxies from the SAM (for model training) and 9,846 galaxies from the COSMOS2025 catalog (for model application).

\subsection{Setting Labels} \label{sec:3:4}
To identify the quiescent population at different epochs in the SAM sample, we adopted the evolving specific star-formation rate (sSFR) threshold from \cite{pacifici2016evolution}, defined as:

\begin{equation}
    \text{sSFR} \leq \text{sSFR}_{lim} = \frac{0.2}{\text{t}_{U}(\text{z})}
\label{eq1}
\end{equation}

Here, sSFR is the specific star-formation rate (in \(\text{Gyr}^{-1}\)), and \(\text{t}_{U}(\text{z})\) is the age of the universe at redshift z (in Gyr). This criterion provides a dynamic threshold that evolves with cosmic time, reflecting the observed decline in galaxy star-formation activity. Applying this criterion to galaxies within the redshift range $2.5 < \text{z} < 5$, we identified 1,870 quiescent galaxies from the 105,876 available galaxies in the SAM sample (the labels are based on the SAM's own true sSFR and redshift).

\subsection{Filling Missing Values} \label{sec:3:5}
The SAM sample contained no missing values in the selected bands (Table~\ref{tab:tab1}). In contrast, the COSMOS2025 sample exhibited missing data in all bands except F444W, which was the band used for sample selection (see Table~\ref{tab:tab2}). To address the missing data in the other ten bands, we employed the \texttt{MissForest} imputation algorithm \citep{stekhoven2015missforest}. This is a robust, non-parametric method that uses an iterative Random Forest model \citep{Breiman2001} to predict missing values.

The algorithm works by modeling each band with missing values as a function of all other bands. In each iteration, a Random Forest is trained on the observed values to predict the missing ones. This process cycles through all affected bands repeatedly, with the imputed values from one step informing the next, until the overall imputation converges and changes minimally. This approach captures complex, non-linear relationships between the photometric bands, ensuring that the imputed magnitudes are statistically consistent with the underlying multi-wavelength structure of the complete data.

This method has been validated on astronomical photometry in our recent study of COSMOS2025 galaxies \citep{AsadiETG_2026}, where it recovered missing values with high accuracy and introduced no systematic bias in color distributions. Given that the missing fractions in the present COSMOS2025 sample are very low (Table~\ref{tab:tab2}, all $\lesssim6\%$), any residual bias from the imputation is expected to be negligible.

\begin{table}
	\centering
	\caption{Percentage of missing values for magnitudes and fluxes in the COSMOS2025 sample across all used bands.}
	\setlength{\belowcaptionskip}{10pt}
	
	\begin{tabular}{c|c|c}
		\hline
		\hline
		COSMOS2025 selected band & \multicolumn{2}{c}{Missing value percentage} \\
		\cmidrule{2-3}
		
		& Magnitude & Flux \\
		\hline
		F115W    & 3.8\%       & 0.0\% \\
		F150W    & 1.3\%       & 0.0\% \\
		F277W    & 0.0\%       & 0.0\% \\
		F444W    & 0.0\%       & 0.0\% \\
		\midrule
		F814W    & 4.1\%       & 0.0\% \\
		\midrule
		Y        & 5.9\%       & 0.0\% \\
		J        & 3.4\%       & 0.0\% \\
		H        & 1.5\%       & 0.0\% \\
		$\text{K}_{\text{s}}$  & 0.9\%       & 0.0\% \\
		\midrule
		ch1      & 1.5\%       & 0.0\% \\
		ch2      & 1.4\%       & 0.0\% \\
		
		\hline
	\end{tabular}
	\label{tab:tab2}
\end{table}

\begin{figure*}
    \centering
    \includegraphics[width=0.7\linewidth]{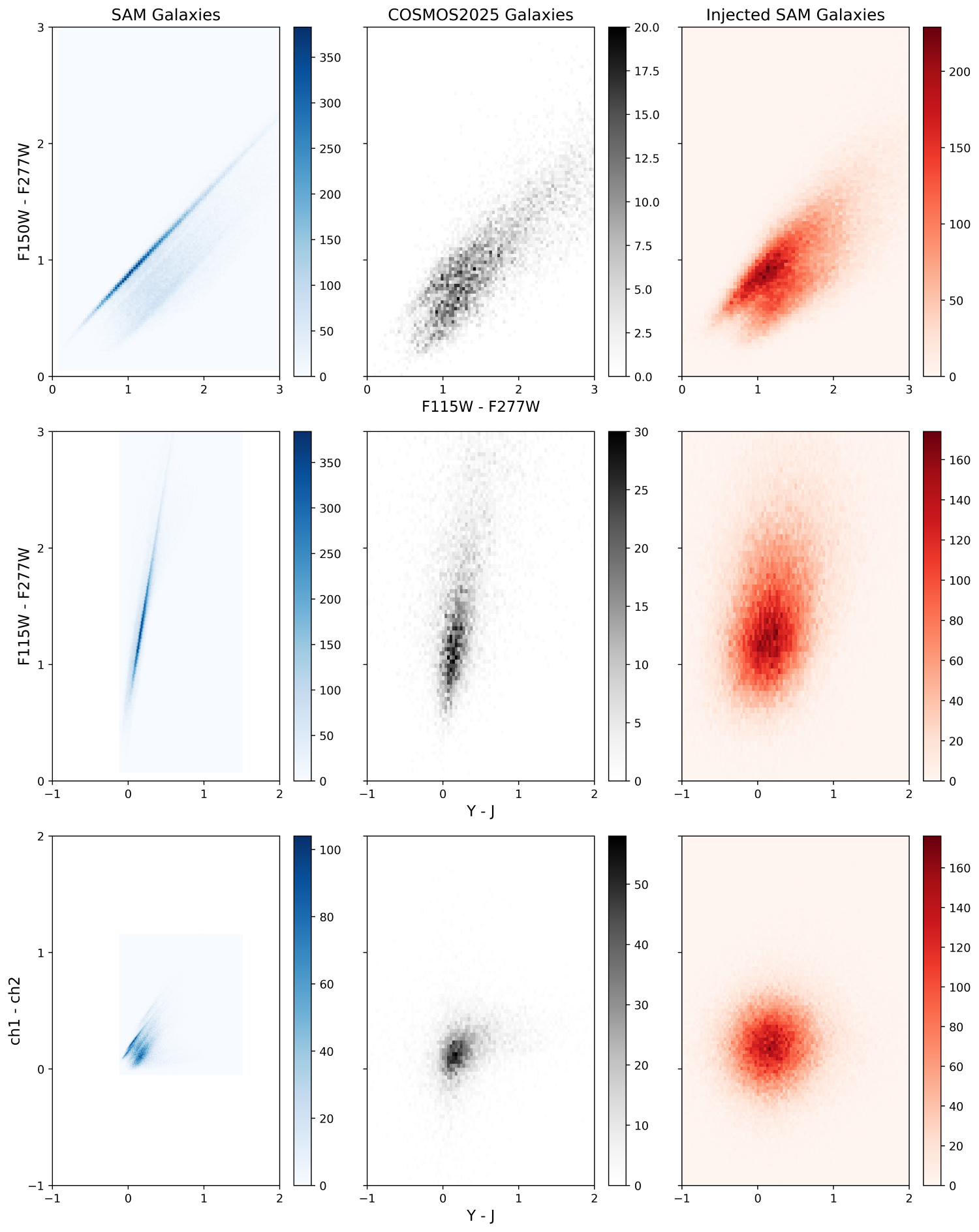}
    \caption{The effect of noise injection on simulated photometry shown through color heatmaps. The left panel shows the original SAM sample, the middle panel shows COSMOS2025 observations, and the right panel shows the SAM sample after noise injection. The noise-injected simulated data closely matches the distribution and scatter of the real observations.}
    \label{fig:fig3}
\end{figure*}

\subsection{Injecting Observational Noise} \label{sec:3:6}
To make the idealized SAM simulated photometry more realistically compatible with the observational COSMOS2025 catalog, we mimic its corresponding color noise properties through a three-step process \citep{asadi_mass}:

\begin{itemize}
\item First, for each of the 55 colors in the COSMOS2025 sample, we trained a \texttt{RandomForestRegressor} model \citep{Breiman2001} using the color values as features and their corresponding color errors as targets.

\item Second, the trained model was applied to predict color errors for the corresponding colors in the SAM sample.

\item Third, to simulate COSMOS-like observational uncertainties, we drew random perturbations $\delta C$ from a zero-centered Gaussian distribution whose standard deviation was equal to the predicted color error. These perturbations were added to the original SAM colors.
\end{itemize}

\begin{figure*}
    \centering
    \includegraphics[width=0.91\linewidth]{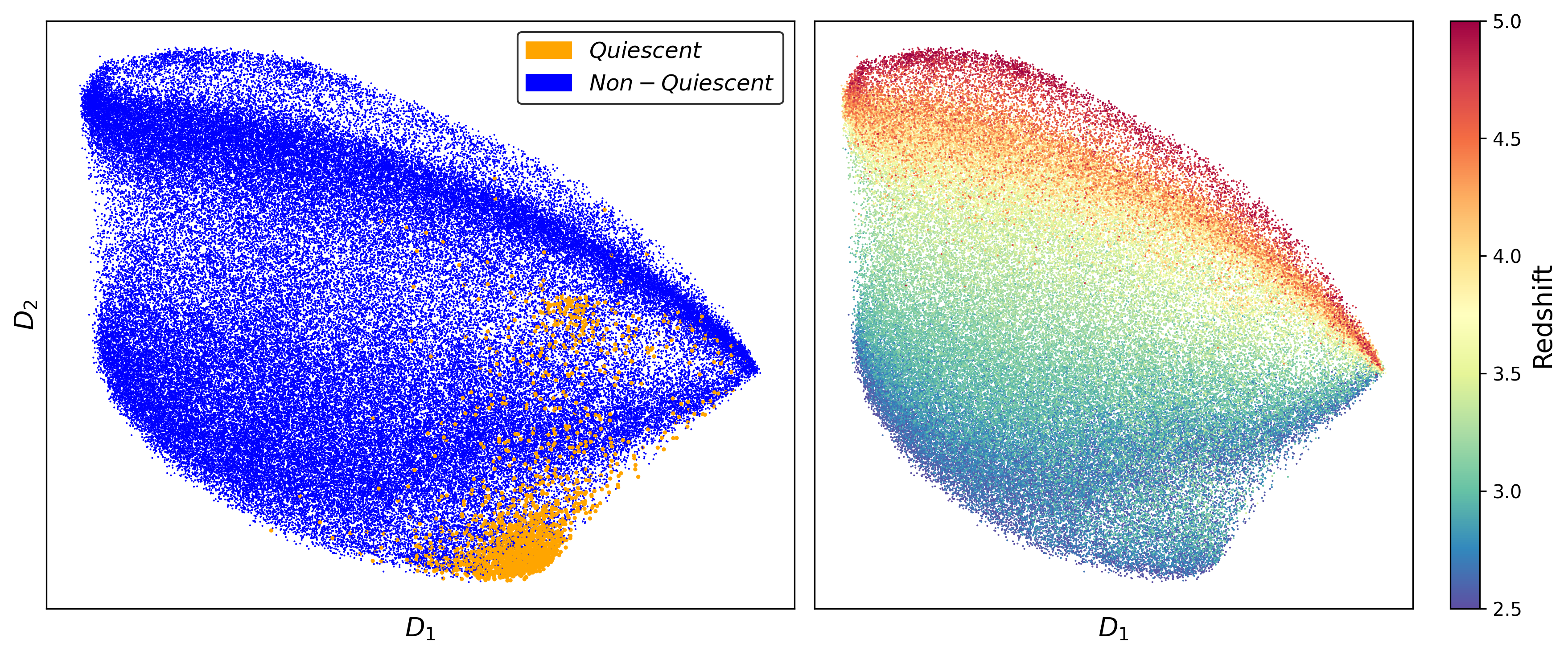}
    \caption{UMAP projection of the SAM sample in color space. Left: Classification labels showing quiescent (orange) and non-quiescent (blue) galaxies. Right: The same projection colored by redshift.}
    \label{fig:fig4}
\end{figure*}

To validate that the noise-injected SAM photometry realistically reproduces the observational scatter, we compared the distributions of some of the colors for the SAM sample (before and after noise injection) and the COSMOS2025 sample. Figure~\ref{fig:fig3} presents this comparison through a triad of heatmaps. The results indicate that the noise-injected SAM galaxies occupy a similar region in color-space as the COSMOS2025 galaxies and effectively cover their data manifold. This suggests that the noise injection process successfully provides a more realistic and representative training set for our ML model.

To provide a quantitative measure of the similarity between the noise-injected SAM and COSMOS2025 color distributions, we performed two-sample Kolmogorov--Smirnov (KS) tests on the colors visualized in Figure~\ref{fig:fig3}. The KS statistic D quantifies the maximum difference between the two cumulative distribution functions, with D = 0 indicating identical distributions. For ch1$-$ch2, Y$-$J, F115W$-$F277W, and F150W$-$F277W, we find D = 0.12, 0.15, 0.15, and 0.12, respectively. Although the associated p-values are formally $<0.001$ due to the large sample sizes ($\sim 10^5$ SAM galaxies and $\sim 10^4$ COSMOS2025 galaxies), the small D values confirm that the noise-injected SAM colors closely reproduce the observed COSMOS2025 color distributions, validating the domain transfer procedure.

To verify that the match holds in the region of color space occupied by the rare quiescent population, we repeat the KS tests using only the quiescent galaxies in the SAM (true sSFR labels) and the ML-quiescent candidates in COSMOS2025 (see Section~\ref{sec:5:3}). For the same four colors, the KS statistics are D = 0.16, 0.18, 0.20, and 0.17, respectively. These values are only slightly larger than those for the full sample, indicating that the noise-injected SAM colors still reproduce the observed quiescent color distributions reasonably well. We note that this comparison is partly self‑referential, as the ML‑quiescent candidates were selected to resemble the SAM quiescent locus; an independent quiescent sample would provide a cleaner test. We therefore describe the match as reasonable rather than precise.

\section{Classification Methods} \label{sec:4}
The core of our analysis involves training a ML model on the noise-injected colors of the SAM sample to learn the classification of quiescent galaxies (Figure~\ref{fig:fig4} shows a two-dimensional Uniform Manifold Approximation
and Projection \citep[UMAP;][]{mcinnes2018umap} of the SAM color space, which separates
the quiescent and non‑quiescent populations and reveals the redshift gradient across the
manifold). The ultimate goal is to apply this trained model to the COSMOS2025 catalog to identify high-redshift quiescent galaxy candidates. However, before proceeding to the observational data, we first evaluate the ML model's performance within the simulation space (the SAM sample) and compare it directly with the parametric SED‑fitting
configuration (exponentially declining SFHs) used for the SAM test set (Section~\ref{sec:4:2}). This comparative analysis in a controlled, simulated environment provides crucial insights and establishes a performance benchmark, which we then use to interpret the model's results in the more complex observational space.

\begin{figure}
    \centering
    \includegraphics[width=\linewidth]{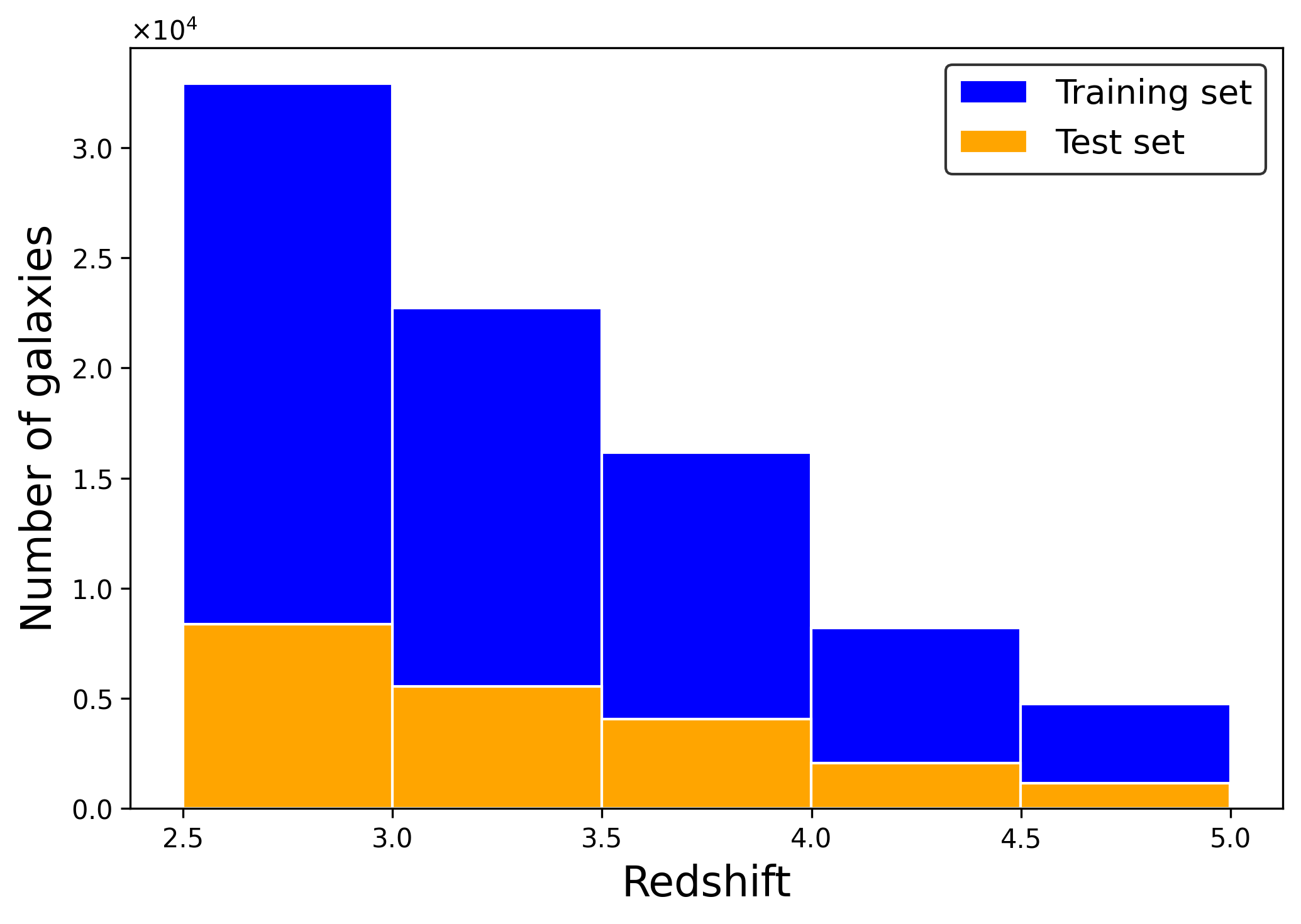}
    \caption{Validation of the training-test split. The redshift distributions for both subsets are shown to be nearly identical, confirming that the random stratified sampling method successfully produced representative and unbiased splits from the SAM sample.}
    \label{fig:fig5}
\end{figure}

To achieve this goal, we first performed a random stratified split of the SAM sample, partitioning it into a training set (80\%) and a test set (20\%). The stratification was based on the labels to ensure that both subsets maintain the same proportion of quiescent and non-quiescent galaxies. This resulted in a training set of 84,700 galaxies (containing 1,496 quiescent galaxies) and a test set of 21,176 galaxies (containing 374 quiescent galaxies). Figure~\ref{fig:fig5} confirms that the redshift distributions between both sets are approximately identical.

\subsection{Machine Learning Method} \label{sec:4:1}
We employed the \texttt{CatBoostClassifier} algorithm \citep{prokhorenkova2018catboost} to classify quiescent galaxies. This method has a strong track record in astronomy; for instance, \cite{Asadi_2025} successfully used it to classify quiescent and star-forming galaxies in the COSMOS2020 catalog. Other applications include distinguishing quasars from galaxies \citep{hughes2022quasar} and identifying Fermi-LAT gamma-ray sources \citep{coronado2022classification}. The algorithm has also been deployed in numerous other astronomical studies \citep[e.g.,][]{humphrey2023euclid, cunha2022photometric, coronado2023redshift, zeraatgari2024exploring, boulet2024catalogue, li2025application}.

\texttt{CatBoostClassifier} is a sophisticated gradient-boosting algorithm designed for classification tasks. It constructs an ensemble of decision trees sequentially, where each successive tree is trained to correct the errors of its predecessors, thereby refining the model's predictions iteratively. This process involves incrementally minimizing a loss function, which enhances the model's ability to generalize from the training data. 

A distinctive feature of the \texttt{CatBoostClassifier} is its implementation of ordered boosting. This is a novel and crucial technique that directly addresses the problem of overfitting, which occurs when a model learns the training data too well and fails to generalize to new, unseen data. By carefully balancing model complexity with the available data, CatBoost's ordered boosting ensures the model remains robust and effective, delivering high performance and stable predictions even on diverse datasets. It achieves this by introducing a more principled and less biased way of calculating gradients, which are the signals the algorithm uses to learn, ultimately resulting in a model that is less prone to memorizing noise in the data \citep[see Section 4.2 in ][ for more details]{prokhorenkova2018catboost}.

While the \texttt{RandomForestClassifier} and other popular ensemble methods like \texttt{XGBoostClassifier} \citep[e.g., ][]{bentejac2021comparative} also utilize an ensemble of trees, they differ in their approach. The \texttt{RandomForestClassifier} builds multiple independent trees on bootstrapped samples of the data and combines their predictions through a voting process. In contrast, both CatBoost and XGBoost are gradient boosting algorithms that build trees sequentially, with each new tree correcting the errors of the previous ones. However, CatBoost's key advantage lies in its novel approach to handling categorical features and its ordered boosting mechanism, which is specifically designed to more effectively prevent overfitting than standard gradient boosting methods like XGBoost.

We trained a \texttt{CatBoostClassifier} using the 55 colors of galaxies in the SAM training set. The model was optimized via a randomized search with 5-fold stratified cross-validation over 50 iterations \citep{pedregosa2011scikit}, aiming to maximize the F1-score (Equation~\ref{eq:5}). The search focused on three key hyperparameters—the number of iterations, learning rate, and tree depth—as they are the primary parameters governing model complexity, convergence, and the fundamental trade-off between bias and variance in gradient-boosting models \citep{geron2019hands}. The optimal values identified through this process are summarized in Table~\ref{tab:tab3}.

Using these optimum values, we retrained the final \texttt{CatBoostClassifier} model on the full training set. This optimized model was then applied to the test set to evaluate its performance, thereby establishing a robust benchmark for the subsequent direct comparison with the SED-fitting method(Section~\ref{sec:4:2}) within the simulation environment.

It is important to mention that prior to any training, we standardized the input features using the \texttt{StandardScaler} \citep{pedregosa2011scikit}, which transforms the data for each color to have a zero mean and unit variance according to:

\begin{equation}
\text{X}_{\text{scaled}} = \frac{\text{X} - \mu}{\sigma},
\label{eq2}
\end{equation}

where $\mu$ is the mean and $\sigma$ is the standard deviation. This pre-processing step ensures that all color features contribute equally to the model's learning process by eliminating biases from their different native scales.

\begin{table}[t]
	\centering
	\caption{Hyperparameter tuning results: search ranges and final selected values.}
	\setlength{\belowcaptionskip}{10pt}
	
	\begin{tabular}{c|ccc}
		\hline
		\hline
		Hyperparameter & Iterations & Learning rate & Depth \\
		\midrule
		Value range & 100--2000 & 0.01--0.3 & 3--10 \\
		\midrule
		Optimum value & 932 & 0.05 & 4  \\
		\hline
		
	\end{tabular}
	\label{tab:tab3}
\end{table}

\subsection{SED-fitting Method} \label{sec:4:2}
To perform a classification based on the SED-fitting method—using the evolving sSFR relation from Equation~\ref{eq1} for a direct comparison with the ML method—we first needed to derive the sSFR values for galaxies in the SAM test set. We obtained these by performing SED-fitting on the set using the \texttt{LePhare} code \citep{arnouts1999measuring, ilbert2006accurate}. We utilized the stellar population synthesis models of \cite{bruzual2003stellar} and assumed an exponentially declining star formation history (SFH), $\mathrm{\text{SFR}} \propto e^{-\text{t}/\tau}$, with nine e-folding timescales ($\tau$) logarithmically spaced from 0.01 to 30 Gyr.

The model grid incorporated a range of physical conditions. Dust attenuation was modeled using the \cite{calzetti2000dust} law with color excess E(B-V) values ranging from 0 to 1. We included contributions from nebular emission lines following \cite{ilbert2008cosmos}. A Chabrier initial mass function \citep{chabrier2003galactic} was adopted over a stellar mass range of 0.01 to $100 \text{M}_{\odot}$, and three metallicities were considered: Z = 0.02, 0.008, and 0.004. The redshift for each galaxy was fixed to the value provided by the SAM catalog.

To validate the SED‑fitting baseline, we compare the \texttt{LePhare}‑derived sSFR and stellar mass against the SAM true values at fixed redshift in the test set. For sSFR, the median offset is $\Delta\log_{10} (\mathrm{\text{sSFR})}=0.14$ dex, the normalized median absolute deviation is $\sigma_{\rm \text{NMAD}}=0.27$ dex, and $15\%$ of galaxies have $|\Delta\log_{10} (\mathrm{\text{sSFR}})| > 0.5$ dex \citep{mobasher2015critical}. For stellar mass, the median offset is $\Delta\log_{10} (\text{M}_*) = 0.02$ dex, $\sigma_{\rm \text{NMAD}} = 0.15$ dex, and $3\%$ of galaxies have $|\Delta\log_{10} (\text{M}_*)| > 0.5$ dex. The redshift was fixed to the SAM truth during the fit, so no redshift validation is performed.

\section{Results} \label{sec:5}
\subsection{Accuracy Metrics} \label{sec:5:1}
To evaluate and compare the performance of our classifiers, we employed standard metrics derived from the confusion matrix \citep{stehman1997selecting}: precision (purity), recall (completeness), and the F1-score. 

\begin{table}[h]
	\caption{The structure of a binary confusion matrix.}
	\centering
	\resizebox{0.45\textwidth}{!}{
		\begin{tabular}{c|c|c}
			& Predicted Positive & Predicted Negative \\
			\hline
			Actual Positive & TP & FN \\
			\hline
			Actual Negative & FP & TN \\
		\end{tabular}
	}
	\label{tab:tab4}
\end{table}

\begin{figure*}
	\centering
	\includegraphics[width=\linewidth]{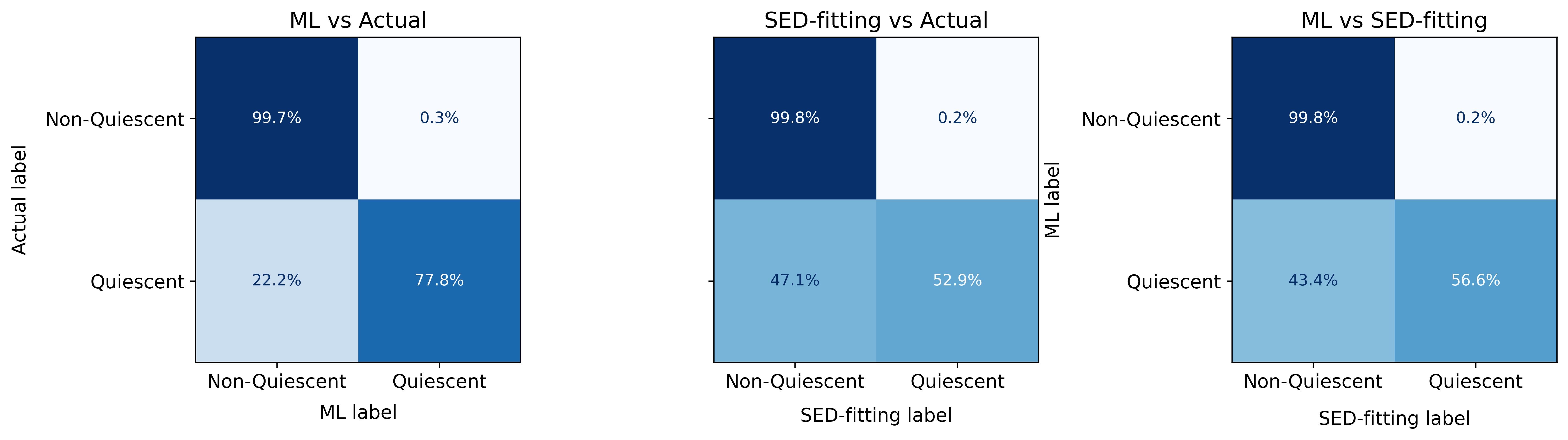}
	\caption{Comparison of classification methods on the SAM test set via confusion matrices. The left panel shows ML predictions versus the actual labels from the mock catalog (Section~\ref{sec:3:4}), the middle panel shows SED-fitting predictions versus the actual labels, and the right panel directly compares predictions between the ML and SED-fitting methods. These matrices quantify agreement and disagreement between classification approaches and ground truth within the simulated test environment.}
	\label{fig:fig7}
\end{figure*}

\begin{table*}[t]
	\centering
	\begin{tabular}{c|c|ccc}
		\hline \hline
		
		Method & Class & Precision & Recall & F1-Score \\
		\midrule
		Machine Learning & Quiescent &$0.82$&$0.78$&$0.80$\\
		& Non-Quiescent &$1.00$&$1.00$&$1.00$\\
		\midrule
		SED-fitting      & Quiescent &$0.85$&$0.53$&$0.65$\\
		& Non-Quiescent &$1.00$&$1.00$&$1.00$\\
		\hline
	\end{tabular}
	\caption{Comparison of the performance of ML and SED-fitting classifiers on the SAM test set of non-quiescent and quiescent galaxies, evaluated using precision (purity), recall (completeness), and F1-score metrics.}
	\label{tab:tab5}
\end{table*}

The confusion matrix is a table used to describe the performance of a classification model on a set of test data for which the true values are known. The matrix itself includes four key terms: True Positives (TP), True Negatives (TN), False Positives (FP), and False Negatives (FN). Table~\ref{tab:tab4} illustrates a confusion matrix for a binary classification.

Precision measures the accuracy of positive predictions, or the purity of the selected positive class.

\begin{equation}
\text{Precision} = \frac{\text{TP}}{\text{TP} + \text{FP}}
\label{eq:3}
\end{equation}

Recall measures the fraction of all actual positives that are correctly identified, or the completeness of the selection.

\begin{equation}
\text{Recall} = \frac{\text{TP}}{\text{TP} + \text{FN}}
\label{eq:4}
\end{equation}

In the specific context of this work, a positive prediction is the identification of a quiescent galaxy. Therefore, a high precision indicates a pure sample of quiescent galaxies, where most of the identified candidates are truly quiescent. A high recall indicates a complete sample, meaning the model successfully recovers a large fraction of the total true quiescent population.

The F1-score, as the harmonic mean of precision and recall, provides a single metric that balances this trade-off between purity and completeness.

\begin{equation}
\text{F1-Score} = 2 \times \frac{\text{Precision} \times \text{Recall}}{\text{Precision} + \text{Recall}}
\label{eq:5}
\end{equation}

\subsection{Method Comparison on SAM Test Set} \label{sec:5:2}
Having established our classification framework, we now evaluate and compare the performance of the ML and SED-fitting methods within the controlled environment of the SAM test set, where the true classification of galaxies is known. This direct comparison is essential for quantifying the relative strengths and weaknesses of each approach before applying them to the observational data.

The confusion matrices for both methods, visualized in Figure~\ref{fig:fig7}, provide the first clear evidence of their differing performance characteristics. For the critical quiescent galaxy class, the SED-fitting method recovers only 53\% of the true quiescent population as TP, while misclassifying a significant 47\% as FN. In stark contrast, the ML classifier achieves a markedly higher TP rate of 78\%, reducing the fraction of missed quiescent galaxies to just 22\%.

A more detailed quantitative comparison is presented in Table~\ref{tab:tab5}, which lists the precision, recall, and F1-score for both methods. The SED-fitting method achieves a high precision (purity) of 85\%, indicating that the galaxies it identifies as quiescent are very likely to be correct. However, this comes at the cost of a low recall (completeness) of 53\%, meaning it fails to identify nearly half of the actual quiescent population. The ML method, while having a slightly lower precision of 82\%, achieves a substantially higher recall of 78\%. The superior balance between these two metrics is captured by the F1-score, which is 80\% for the ML method compared to 65\% for SED-fitting.

\begin{figure}
	\centering
	\includegraphics[width=\linewidth]{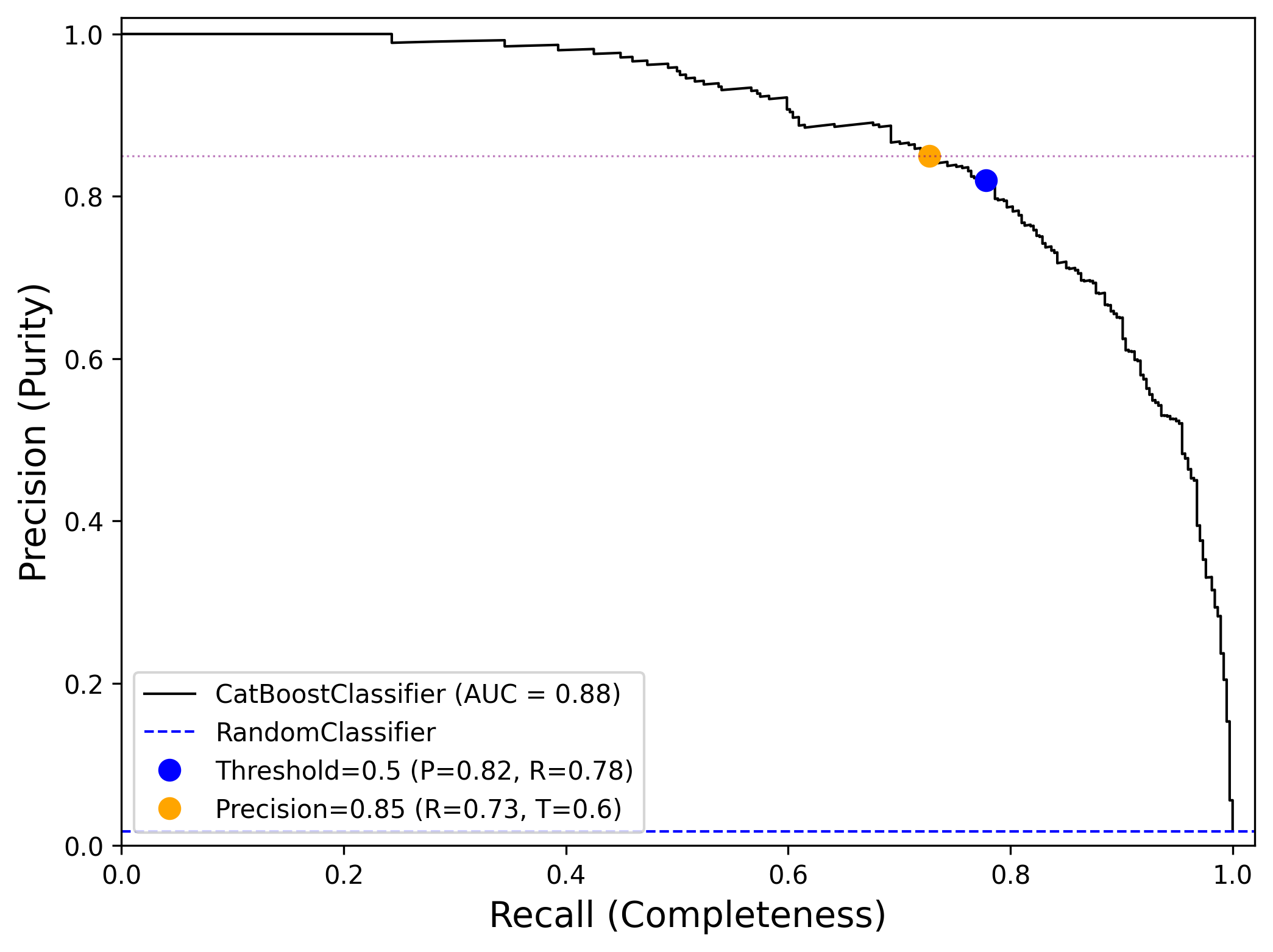}
	\caption{Precision-recall curve for the ML method in identifying quiescent galaxies within the SAM test set, achieving an area under the curve (AUC) of 0.88. The blue circle indicates the default operating point (probability threshold = 0.5), yielding a precision of 82\% and a recall of 78\%. The orange circle marks an alternative, more conservative threshold (0.6) that matches the SED-fitting method's precision of 85\%, while maintaining a superior recall of 73\%. The blue dashed line represents the performance of a random classifier.}
	\label{fig:fig8}
\end{figure}

The trade-off between purity and completeness in the ML classifier is further illuminated by its precision-recall curve (Figure~\ref{fig:fig8}), which has an Area Under the Curve (AUC) of 0.88, indicating strong overall performance. The default classification threshold of 0.5 was chosen to provide a good balance, yielding the performance metrics in Table~\ref{tab:tab5}. It is noteworthy that if we were to adjust the ML classification threshold to match the SED-fitting method's precision of 85\%, we would require a threshold of approximately 0.6. At this more conservative threshold, the ML classifier would still maintain a recall of 73\%. This demonstrates that even when matched in purity, the ML method's completeness is 20 percentage points (or $\sim$38\%) higher than that of the SED-fitting method.

This key result underscores a fundamental difference: the SED-fitting method, while pure, misses a significant fraction of the quiescent population, whereas the ML method constructs a more complete census of quiescent galaxies with only a marginal reduction in purity.

\begin{figure*}
	\centering
	\includegraphics[width=\linewidth]{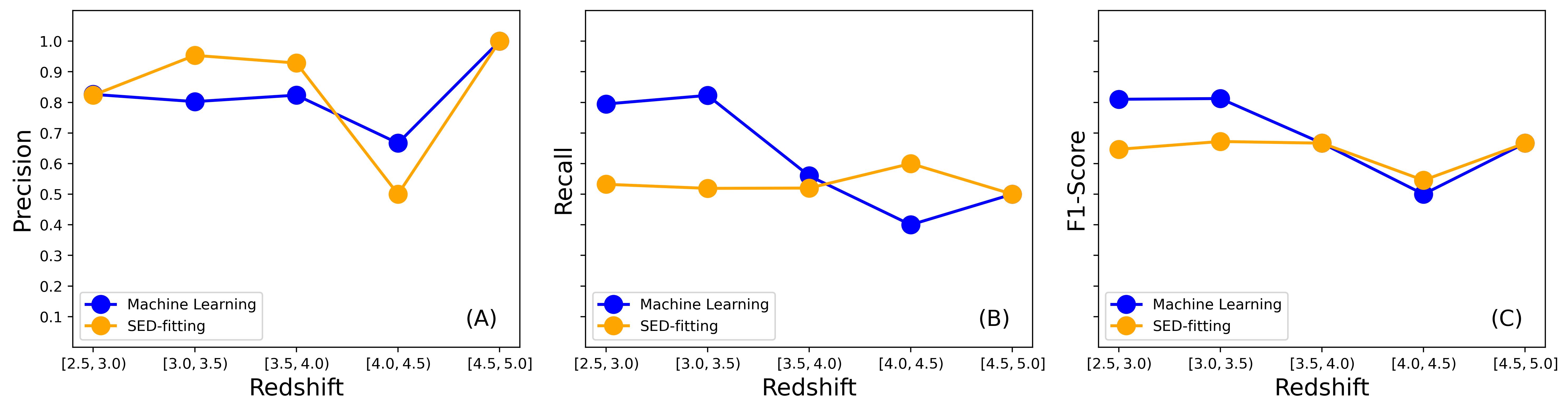}
	\caption{Redshift-resolved performance comparison of the ML and SED-fitting methods for identifying quiescent galaxies in the SAM test set. Panels show (A) precision (purity), (B) recall (completeness), and (C) F1-score for both classifiers across five redshift bins spanning $2.5 < \text{z} < 5.0$.}
	\label{fig:fig9}
\end{figure*}

While the global metrics demonstrate the overall superiority of the ML approach, a more nuanced picture emerges when we examine performance as a function of redshift. Figure~\ref{fig:fig9} presents the precision, recall, and F1-score for both classification methods across five redshift bins within the $2.5 < \text{z} < 5$ range, providing critical insight into how their performance evolves towards the epoch of early galaxy formation.

\begin{figure*}
	\centering
	\includegraphics[width=0.91\linewidth]{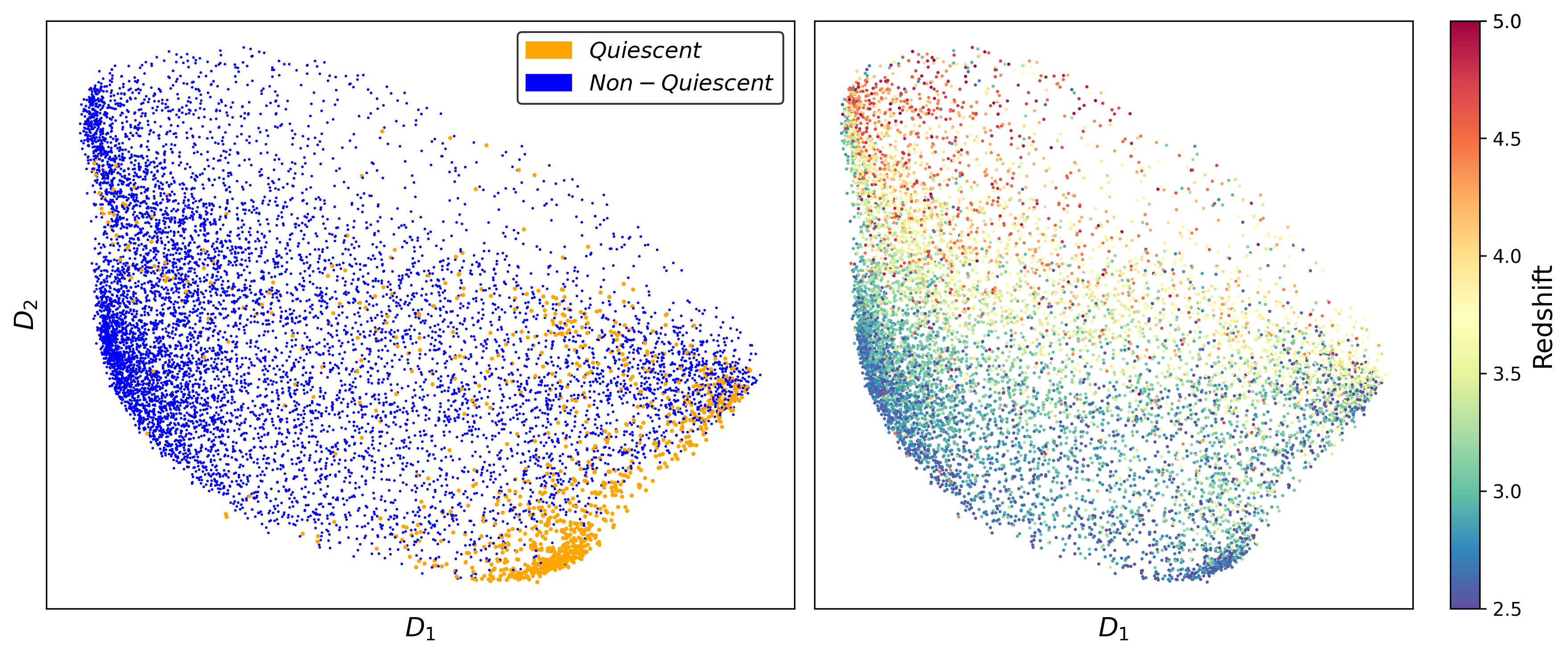}
	\caption{UMAP projection of the COSMOS2025 sample in color space. Left: ML Classification labels showing quiescent (orange) and non-quiescent (blue) galaxies. Right: The same projection colored by redshift.}
	\label{fig:fig10}
\end{figure*}

The redshift-dependent performance reveals several key trends. As shown in panel (A), both methods maintain high precision ($\gtrsim$80\%) across most bins, with SED-fitting occasionally achieving near-perfect purity (e.g., 95\% in the $3.0 \leq \text{z} < 3.5$ bin). However, panel (B) unveils the fundamental weakness of the SED-fitting approach: its recall remains consistently and significantly lower than the ML method across all redshifts. The ML classifier sustains a recall above 79\% in the two lowest redshift bins, only declining to 40\%-50\% at $\text{z} > 4.0$. In stark contrast, the SED-fitting method struggles to recover even 60\% of the true quiescent population at any redshift, with its recall falling to approximately 53\% in the lower-redshift bins where these galaxies are most abundant. This completeness gap directly translates to the superior F1-scores of the ML method, shown in panel (C), which consistently outperforms SED-fitting by $\Delta \text{F1-score} \approx 0.14$-$0.16$ across the first three bins. Notably, in the highest redshift bin ($4.5 \leq \text{z} \leq 5.0$), both methods converge due to the extremely small sample statistics, though the ML classifier still achieves perfect precision while matching SED-fitting's recall.

\subsection{ML Classification of COSMOS2025 Galaxies} \label{sec:5:3}
Having demonstrated the superior performance, particularly in completeness, of the ML classifier over SED-fitting within the simulated SAM test environment, we now apply the trained model to the COSMOS2025 observational catalog.

We applied the final \texttt{CatBoostClassifier}, trained on the entire SAM sample (combining the previous training and test sets), to the pre-processed COSMOS2025 sample of 9,846 galaxies described in Section~\ref{sec:3:3}. Prior to classification, we constructed the identical set of 55 color indices from the COSMOS2025 photometry and applied the same standardization procedure (Equation~\ref{eq2}) using the mean and standard deviation values derived from the SAM sample data.

To visualize the distribution of the COSMOS2025 sample in the multi-dimensional color space, Figure~\ref{fig:fig10} presents a UMAP projection analogous to Figure~\ref{fig:fig4} for the SAM sample. This projection reveals the underlying structure of the observational data and provides context for where the ML-classified quiescent galaxies reside within the color manifold.

\begin{table*}
	\centering
	\caption{Comparison of Quiescent Galaxy Populations Identified by Different Methods in COSMOS2025 sample.}
	\begin{tabular}{c|c|ccccc}
		\hline
		\hline
		Method & Total & $2.5 \leq \text{z} < 3.0$ & $3.0 \leq \text{z} < 3.5$ & $3.5 \leq \text{z} < 4.0$ & $4.0 \leq \text{z} < 4.5$ & $4.5 \leq \text{z} \leq 5.0$ \\
		\hline
		Machine Learning & 1111 & 416 & 391 & 215 & 61 & 28 \\
		SED-fitting & 427 & 159 & 127 & 75 & 41 & 25 \\
		\hline
		
	\end{tabular}

	\label{tab:tab6}
\end{table*}

The classifier identified 1,111 galaxies as quiescent, representing approximately 11.3\% of the massive galaxy population in this redshift range. Table~\ref{tab:tab6} presents the distribution of these quiescent candidates across redshift bins, revealing a clear evolutionary trend where the number of quiescent galaxies decreases with increasing redshift, as expected from cosmological downsizing.

\begin{figure}
	\centering
	\includegraphics[width=\linewidth]{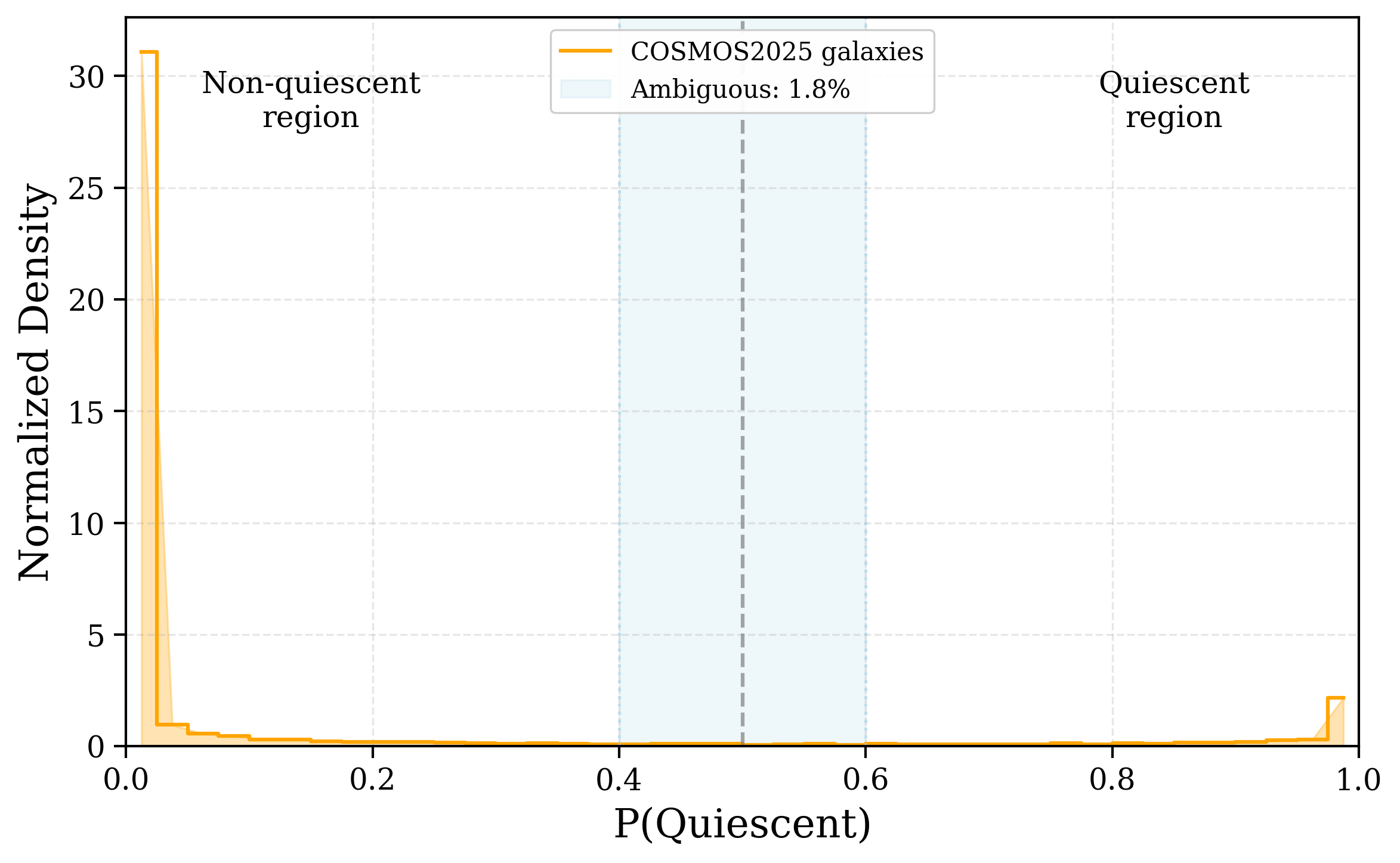}
	\caption{Distribution of predicted quiescent probabilities for COSMOS2025 galaxies. Quiescent candidates and non-quiescent galaxies are strongly separated. Only 1.8\% of the sample falls in the ambiguous $0.4 < \text{p} < 0.6$ range, demonstrating the classification's robustness to threshold variation.}
	\label{fig:prob_hist}
\end{figure}

To assess the robustness of these classifications, we examine the distribution of predicted quiescent probabilities output by the \texttt{CatBoostClassifier}. Figure~\ref{fig:prob_hist} displays the normalized histogram of probabilities for all COSMOS2025 galaxies in our sample, separated by the final class assignment at the 0.5 threshold. The distribution is strongly bimodal: non-quiescent galaxies cluster at probabilities near zero, while the quiescent candidates predominantly lie near unity. Only 1.8\% of the sample (175 galaxies) fall within the intermediate range $0.4 < \text{p} < 0.6$, and of these, merely 19 are classified as quiescent. This clear separation demonstrates that the classification is not sensitive to the precise choice of the decision threshold; varying the threshold by $\pm 0.1$ changes the quiescent candidate count by less than 2\%. This small change in the total candidate count is consistent with the decline in recall observed on the SAM test set (from 78\% to 73\% for the same threshold shift; Figure~\ref{fig:fig8}), because the recall measures the fraction of true quiescent galaxies recovered, whereas the candidate count reflects the total number of classified objects in the COSMOS2025 sample, where the vast majority of galaxies lie far from the decision boundary. The high confidence of the quiescent sample reinforces the reliability of the 1,111 candidates reported here.

To assess the sensitivity of our results to the adopted selection boundaries, we varied the redshift limits by $\pm 0.1$ and the stellar mass limit by $0.2$ dex, chosen to be comparable to the typical uncertainties in photometric redshifts and SED-fitted masses in our redshift and mass range. The total number of ML quiescent candidates changes by at most $\sim10.3\%$, and the ML--to--SED candidate ratio remains $2.6\pm 0.1$ across all variations. This suggests that the main conclusions are not strongly sensitive to modest selection-level mismatches between the SAM (true quantities) and COSMOS2025 (estimated quantities). We note that this test varies the selection limits coherently and does not simulate per‑object photometric‑redshift scatter or catastrophic photo--z failures. A full perturbation of the mock true redshifts with a realistic error‑plus‑outlier distribution would provide a more direct assessment; the mass validation in Section~\ref{sec:4:2} indicates that the stellar‑mass leg of the cut is minor, so the residual uncertainty is dominated by photo--z scatter.

\subsection{Method Comparison on COSMOS2025 Galaxies} \label{sec:5:4}
We now compare the population of quiescent galaxies identified by our ML classifier with those selected using the catalog’s parametric SED‑fitting approach in the COSMOS2025 sample. Applying the same evolving sSFR threshold (Equation~\ref{eq1}) to the \texttt{LePhare}-derived sSFR values from the catalog (derived from a library of exponentially declining and delayed SFHs), we identify 427 galaxies as quiescent via SED-fitting. This represents only 38\% of the 1,111 candidates identified by the ML method, highlighting a substantial discrepancy in the census of the high-redshift quiescent population between the two approaches.

Table~\ref{tab:tab6} presents the redshift distribution of quiescent galaxies for both methods, revealing consistent trends but dramatically different absolute numbers. While both methods show the expected decrease in quiescent galaxy counts with increasing redshift, the ML method identifies 2.6 times more quiescent galaxies overall. This difference is most pronounced at lower redshifts ($2.5 \leq \text{z} < 3.5$), where the ML method recovers approximately 2.5-3 times more candidates.

The agreement between the two methods shows high precision (91\%) but low recall (35\%) when treating the ML classifications as ground truth and SED-fitting as predictions. This pattern indicates that when the two methods agree on a galaxy being quiescent, they are highly consistent (high purity), but SED-fitting fails to identify the majority of galaxies that the ML method classifies as quiescent (low completeness). This pattern of high purity but poor completeness for the SED-fitting method is fully consistent with our findings from the SAM test set, where the simpler parametric SED‑fitting configuration showed precision = 85\% and recall = 53\% with respect to the ML classifications labels.

We note that the COSMOS2025 catalog's \texttt{LePhare} configuration is more flexible than the one adopted for the SAM test set: it includes both exponentially declining and delayed SFHs, three dust attenuation curves, and two metallicities \citep{shuntov2025cosmos2025}. Even with this increased flexibility, the catalog's SED-fitting yields 2.6 times fewer quiescent candidates than the ML classifier. This comparison therefore shows a completeness gap for the specific \texttt{LePhare} configurations examined here, but it does not by itself rule out the possibility that a more flexible forward-modeling or non-parametric SED-fitting approach could recover a larger fraction of the quiescent population.

\subsection{Properties of ML-Only Quiescent Candidates}
\label{sec:5:5}
\begin{figure}
	\centering
	\includegraphics[width=\linewidth]{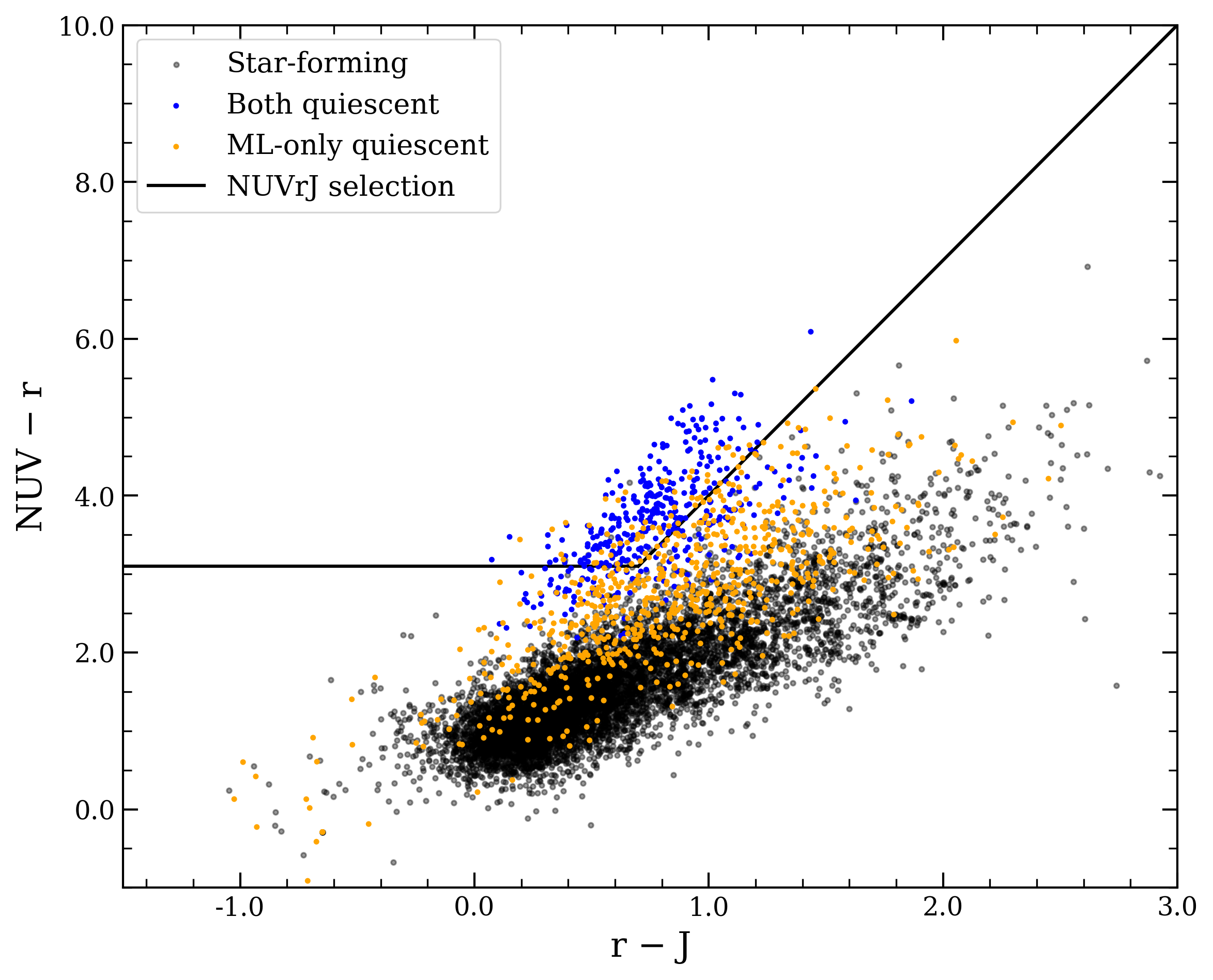}
	\caption{
		Rest-frame NUV$-$r vs.\ r$-$J diagram for COSMOS2025 galaxies sample. Black points: star-forming galaxies (ML-classified as non-quiescent). Blue points: galaxies classified as quiescent by both the ML model and SED-fitting. Orange points with density contours: ML-only quiescent candidates (ML-quiescent but SED-non-quiescent). The solid black line marks the NUVrJ quiescent selection boundary.}
	\label{fig:nuvrj}
\end{figure}

The 684 galaxies identified as quiescent by the ML classifier but not by the SED-fitting criterion (hereafter ML-only) constitute the most distinctive subset of our sample. To understand their nature, we examine their rest-frame NUV$-$r and r$-$J colors. Figure~\ref{fig:nuvrj} presents the NUVrJ diagram for the COSMOS2025 sample. The solid black line marks the canonical quiescent selection boundary, defined by $\text{NUV}-\text{r} > 3.1$ and $\text{NUV}-\text{r} > 3\times(\text{r}-\text{J}) + 1$. Galaxies classified as quiescent by both methods (blue points) lie predominantly within this wedge, with median colors r$-$J = 0.77 and NUV$-$r = 3.68. In contrast, the ML-only candidates (orange points) exhibit systematically bluer NUV$-$r (median 2.76) while maintaining comparably red r$-$J (median 0.85). The median vertical offset of the ML-only population below the selection boundary is 1.14 mag. Approximately one-third of ML-only candidates fall within 1.0 mag of the threshold, while the majority occupy a transitional zone between the fully quiescent and actively star-forming loci.

We note that the SAM validation (Section~\ref{sec:5:2}) indicates an expected purity of $\sim$82\% for the ML classifier. Assuming this purity translates to the COSMOS2025 sample, approximately 18\% of the 684 ML-only candidates may be false positives (i.e., star-forming galaxies misclassified as quiescent). The remainder likely represent genuine transitional or recently quenched systems, whose colors are consistent with post-starburst galaxies.

\section{Discussion} \label{sec:6}
This study has presented a ML framework to identify massive quiescent galaxies at high redshifts in the COSMOS2025 catalog. By training a sophisticated classifier on physically-motivated simulations from the Santa Cruz SAMs—incorporating key JWST bands and realistic observational effects—we developed a model for distinguishing quiescent galaxies based on their multi-wavelength colors. Our validation within the simulated environment demonstrated that this ML approach offers a significant improvement in completeness over the parametric SED‑fitting configuration tested here while maintaining high purity. When applied to the observational data, the classifier uncovered a significantly larger population of quiescent candidates than the catalog’s parametric SED‑fitting method. This consistent pattern between simulation and observation underscores the capability of ML methods to reveal a more complete census of the early quiescent population.

\subsection{Interpretive Framework: The SAM Ground Truth} \label{sec:6.1}
A foundational consideration for interpreting the results of this study is the dependence of the ML classifier on the specific definition of quiescence encoded in the Santa Cruz SAMs. The model is optimized to reproduce the evolving sSFR threshold of Equation~\ref{eq1} as applied to the stochastic SFHs of the simulation. Consequently, the performance comparison presented in Section~\ref{sec:5:2} demonstrates that the ML classifier is highly effective at recovering SAM-defined quiescent galaxies, while the specific SED-fitting configuration tested here is less complete in reproducing the same labels. The quiescent candidates identified in COSMOS2025 therefore represent galaxies whose colors are consistent with the SAM's specific quenching pathways. If the physical processes governing quenching in the real Universe differ from those implemented in the SAM, the ML model may over- or under-estimate the true quiescent fraction. Conversely, if the SAM provides a reasonable approximation of high-redshift galaxy evolution, the ML method offers a meaningful improvement in completeness over template fitting.

We note that the NUVrJ comparison in Section~\ref{sec:2:1} serves a different purpose from the \texttt{LePhare}-based classification. The agreement between the SAM sSFR-based quiescent fraction and the observed NUVrJ fraction (Figure~\ref{fig:fig1}) indicates that the SAM reproduces the broad rest-frame color locus of the observed galaxy population, which is a necessary condition for any color-based selection. It does not, however, validate NUVrJ as a complete quiescent selector. The \texttt{LePhare} comparison tests the completeness of a specific template-fitting and sSFR-threshold procedure; the fact that it finds fewer candidates than the ML classifier highlights the incompleteness of that particular method, while not, by itself, undermining the SAM realism check. The two checks are therefore complementary: one assesses the overall realism of the mock colors, while the other probes the sensitivity of a derived-quantity classification to the assumed SFH.

\subsection{Reasons for ML's Higher Fidelity in Recovering SAM Labels} \label{sec:6.2}
Given the conditional nature of the comparison established above, it remains instructive to understand why the ML classifier recovers the SAM ground truth more completely than the simple SED‑fitting configuration (exponentially declining SFHs) used for the SAM test set in Section~\ref{sec:4:2}. This performance difference can be attributed to several key factors rooted in how each method translates photometric data into a classification.

First, the SED-fitting process is inherently dependent on the assumed model templates and SFHs. These templates (e.g., exponentially declining models) are often too simple for the complex, stochastic SFHs present in physically-motivated SAMs. This modeling mismatch is likely a major contributor of the SED-fitting method's poor recall (the sSFR validation in
Section~\ref{sec:4:2} shows that a median offset of 0.14 dex and a scatter of $\sigma_{\rm \text{NMAD}}=0.27$ dex scatter galaxies near the quiescence threshold, contributing to the 53\% recall. But we note that this analysis does not distinguish between genuine SFH-model mismatch and simple estimation scatter in the derived sSFR; disentangling these two sources of incompleteness would require a dedicated forward‑modeling experiment.) as it leads to misclassification of galaxies in transitional phases, whose SEDs are not well-captured by simple parametric histories. For instance, recently quenched galaxies, where A-type stars still contribute significantly to the SED, or young, dusty galaxies whose colors can mimic those of old, quiescent systems, present a well-known challenge for template-fitting methods. While the accuracy of SED-fitting can be improved with more sophisticated codes—such as those employing Bayesian inference \citep{leja2019measure, johnson2021stellar, hahn2022accelerated} or non-parametric SFHs—this typically comes at a steep computational cost as the model parameter space expands dramatically.

Second, while the overall analysis uses photometric redshifts and SED‑fitted stellar masses to select the COSMOS2025 sample, the ML classifier itself does not need to estimate sSFR from photometry. Instead of relying on a potentially imperfect inversion from observed light to physical parameters, the classifier directly learns the complex mapping between a galaxy's multi‑color signature and its SAM‑defined label. This reduces the propagation of errors that can arise from the
simplifying assumptions inherent in parametric SED‑fitting models.

Third, the ML model leverages the full dimensionality of the photometric data in a holistic manner. Unlike SED-fitting, which effectively fits each galaxy's SED in isolation, the ML algorithm learns the relative discriminatory power of all 55 colors simultaneously across the entire training population. Colors that are less informative for the classification task are automatically down-weighted, allowing the model to focus on the most salient features in the data.

Finally, the two methods are optimized for different objectives. The SED-fitting method is primarily designed to estimate physical parameters, with classification being a secondary step based on a derived quantity (sSFR). In contrast, the ML model is trained end-to-end with the explicit and singular goal of maximizing classification accuracy. This focused objective function allows it to find a more effective decision boundary in the high-dimensional color space, even if the intermediate ``reasoning" is less interpretable than a physically derived sSFR.

\subsection{Scientific Implications Under the SAM Framework} \label{sec:6.3}
The most significant outcome of our analysis is the substantially larger population of quiescent galaxies identified by the ML method compared to the catalog’s parametric SED‑fitting approach. Under the working assumption that the SAM definition of quiescence is a reasonable proxy for real galaxy evolution, this discrepancy suggests that such simple parametric SED‑fitting techniques may miss a specific sub-population, likely objects in the early stages of quenching whose SEDs are not well-captured by simple parametric SFHs (the rest-frame NUVrJ colors of the ML-only candidates in Section~\ref{sec:5:5} are consistent with this interpretation). If confirmed, this larger census would imply that the mechanisms responsible for quenching star formation at high redshifts may need to be more efficient than currently inferred from SED-fitting studies alone.

\subsection{Limitations and Future Work} \label{sec:6.4}
Beyond the foundational SAM-dependency discussed in Section~\ref{sec:6.1}, several additional limitations warrant mention. 

As discussed in Section~\ref{sec:6.2}, the SED-fitting comparison employs a relatively simple parametric configuration (exponentially declining SFHs). More flexible non-parametric SFH codes would likely recover a higher fraction of recently quenched galaxies, and future work should compare against such methods.

While our pre-processing pipeline—including the injection of observational noise and the careful matching of photometric bands—was explicitly designed to bridge this gap and enhance the model's transferability, a residual domain shift may persist. We also note that the noise-injection procedure assumes uncorrelated, Gaussian photometric errors, whereas real uncertainties can be covariant and have non‑Gaussian tails. The KS tests (Section~\ref{sec:3:6}) suggest that this simplification does not significantly affect our results, although a fully self-consistent treatment with realistic correlated noise would be valuable.

The true robustness of our approach would be further strengthened by training on spectroscopically confirmed samples of high-redshift quiescent galaxies, though such samples are currently small. Future work could also validate the model against mock catalogs from diverse simulations (e.g., hydrodynamical models) to ensure the findings are not unduly influenced by the specific implementation of one particular semi-analytic model.

Finally, a quantitative assessment of the impact of this larger quiescent candidate sample on galaxy evolution models will require future computation of comoving number densities and stellar mass functions.

\section{Conclusion} \label{sec:7}
In this study, we have developed a machine learning framework to conduct a census of massive quiescent galaxies at $2.5 < \text{z} < 5$ within the COSMOS2025 catalog. The methodology involved training a \texttt{CatBoostClassifier} on noise-injected mock observations from the Santa Cruz SAMs, leveraging 55 color features derived from key JWST NIRCam and ancillary bands. This approach offers a new, more complete perspective on the early quiescent population. Our principal conclusions are as follows:

\begin{itemize}
	\item In a controlled test against known truths within the SAM environment, our ML classifier demonstrates higher fidelity than the parametric SED‑fitting configuration tested here in recovering the SAM-defined quiescent labels. It achieved a significantly higher recall (completeness) of 78\%, compared to 53\% for SED-fitting, while maintaining a high purity of 82\%.
	
	\item Applying the trained model to the COSMOS2025 catalog--under the assumption that the SAM definition of quiescence transfers to the real Universe--we identified 1,111 massive quiescent galaxy candidates, a population 2.6 times larger than the 427 candidates identified through the simple parametric SED‑fitting used in the catalog.
	
\end{itemize}

The consistent pattern of high purity but poor completeness for the tested SED‑fitting configurations, observed across both the simulated test set and the COSMOS2025 application, suggests that the specific \texttt{LePhare} configuration adopted here is less complete than the ML classifier in recovering quiescent galaxies as defined by the SAM framework. This discrepancy indicates that, under the assumptions of our analysis, the tested SED-fitting baseline may miss part of the quiescent population. Within this context, our ML approach provides a more complete census of massive quiescent galaxy candidates at high redshift.

\section*{Acknowledgments}
We thank the anonymous referee for their thorough and constructive comments, which have significantly strengthened the methodological rigor and clarity of this manuscript.

\section*{Data Availability}
The trained ML model and full classified COSMOS2025 sample 
(including quiescent and non‑quiescent galaxy classifications) resulting from this analysis are publicly available on Zenodo at \url{https://doi.org/10.5281/zenodo.19830179}, and also in the GitHub repository: 
\url{https://github.com/vahidoo7/ML-High-redshift-Quiescent-Galaxy-Identifier}.

\software{Pandas \citep{mckinney2011pandas}, Scikit-learn \citep{pedregosa2011scikit}, Numpy \citep{harris2020array}, UMAP \citep{mcinnes2018umap}, Catboost \citep{prokhorenkova2018catboost}, Astropy \citep{robitaille2013astropy, astropy_2018, astropy_2022}, Seaborn \citep{waskom2021seaborn}, Matplotlib \citep{barrett2005matplotlib}, SciPy \citep{virtanen2020scipy}, MissForest \citep{stekhoven2015missforest}, LePhare \citep{arnouts1999measuring}.}


\bibliography{ref}

\end{document}